\input{aipcheck}

\documentclass[
    ,final            % use final for the camera ready runs
  ]
  {aipproc}

\layoutstyle{6x9}
\usepackage{graphics}

\renewcommand\XFMtitleblock{
  \XFMtitle
  \let\XFMoldpar\par
  \def\par{\XFMoldpar\def\par{\space
        for the CTA consortium\XFMoldpar}}
   \XFMauthors
   \let\par\XFMoldpar
   \XFMaddresses
   \XFMabstract
   \vspace{5pt}
   \XFMkeywords
   \XFMclassification
}
\newcommand{\gray}{$\gamma$-ray}

\begin{document}

\title{Advanced analysis and event reconstruction for the CTA Observatory}

%95.85.Pw 	γ-ray 
%95.75.-z 	Observation and data reduction techniques; computer modeling and simulation
%95.75.Pq 	Mathematical procedures and computer techniques 

\author{Y. Becherini}{
  address={Astroparticule et Cosmologie, CNRS/Univ.\ Paris 7, B\^at.\ Condorcet, Paris 75205 cedex 13, France},
  altaddress={Laboratoire Leprince-Ringuet , Ecole Polytechnique, CNRS/IN2P3, F-91128 Palaiseau, France}
}

\author{B. Kh\'elifi}{
  address={Laboratoire Leprince-Ringuet , Ecole Polytechnique, CNRS/IN2P3, F-91128 Palaiseau, France}
}

\author{S. Pita}{
  address={Astroparticule et Cosmologie, CNRS/Univ.\ Paris 7, B\^at.\ Condorcet, Paris 75205 cedex 13, France}
}

\author{M. Punch}{
  address={Astroparticule et Cosmologie, CNRS/Univ.\ Paris 7, B\^at.\ Condorcet, Paris 75205 cedex 13, France}
}

\begin{abstract}

The planned Cherenkov Telescope Array (CTA) is a future observatory for very-high-energy (VHE) gamma-ray astronomy composed of one site per hemisphere \cite{DesignConcepts}. It aims at 10 times better sensitivity, a better angular resolution and wider energy coverage than current installations such as H.E.S.S., MAGIC and VERITAS.

In order to achieve this level of performance, both the design of the telescopes and the analysis algorithms are being studied and optimized within the CTA Monte-Carlo working group. Here, we present ongoing work on the data analysis for both the event reconstruction (energy, direction) and gamma/hadron separation, carried out within the HAP (H.E.S.S. Analysis Package) software framework of the H.E.S.S. collaboration, for this initial study. The event reconstruction uses both Hillas-parameter-based algorithms and an improved version of the 3D-Model algorithm \citep{model3D}. For the gamma/hadron discrimination, original and robust discriminant variables are used and treated with Boosted Decision Trees (BDTs) in the TMVA \citep{TMVA} (Toolkit for Multivariate Data Analysis) framework. With this advanced analysis, known as Paris-MVA \citep{ParisMVA}, the sensitivity is improved by a factor of $\sim 2$ in the core range of CTA relative to the standard analyses.

Here we present the algorithms used for the reconstruction and discrimination, together with the resulting performance characteristics, with good confidence, since the method has been successfully applied for H.E.S.S.

\end{abstract}

\maketitle

%%%%%%%%%%%%%%%%%%%%%%%%%%%%%%%%%%%%%%%%%%%%
%% MAINMATTER
%%%%%%%%%%%%%%%%%%%%%%%%%%%%%%%%%%%%%%%%%%%%

\section{Paris-MVA Advanced Analysis}

\subsection{General Description}

The Paris-MVA advanced analysis method is based on a combination of both already-known and newly-derived parameters, combined in an Multi-Variate analysis (MVA). Details can be found in \citep{ParisMVA} \& references therein.

A total of eight parameters are used as input. The first two are standard ``Hillas'' image moment-based parameters: Mean-scaled width and Mean-scaled length.
The 3D-Model \citep{model3D} provides three further event parameters which are powerful discriminants: 
3D-Reduced-Width, its error as given by the likelihood maximization, and the fitted Depth of Shower Maximum.
Finally three newly-defined parameters have been developed, which are described below and in \citep{ParisMVA}.
All these parameters are input to the ROOT software suite's TMVA framework \citep{TMVA}, applying an MVA based on Boosted Decision Trees (BDTs). 
Careful testing of the decision tree design (depth, pruning etc.) was carried out, for example to avoid over-training.  
In addition, it should be noted that the input parameters were chosen from a larger set using the tests provided in the TMVA package to avoid correlated parameters, 
since these reduce the efficiency of the BDT procedure.

%\begin{figure}
%  \includegraphics[width=16cm]{Figure_ParamDist}
%  \caption{Distributions of the ($\log10$ of) 3D-model discriminant parameters, in one example training bin from CTA simulations.  Signal is shown in blue histograms, Background in %red.}
%  \label{ParamDist}
%\end{figure}

The training of the BDTs was carried out in a large number of bins (27): over a wide range in Energy, for nine bins in a related 3D-model parameter, $N_{\rm phot}$, which provides an estimate of the number of Cherenkov photons emitted by the shower; %'s photosphere, see below; 
and for three classes in event multiplicity (2, 3--5, and 6--60 telescopes events) independently of telescope type.
%For the simulations used, only a single Zenith angle was available, but in future work several bins in Zenith angle will be used to take into account real observation conditions.

The BDT procedure combines the parameters for each bin in an optimal manner, to provide a single output BDT-parameter on which to cut.
Note that the relative efficiency of different parameters varies widely, especially as a function of energy.
For Paris-MVA, we choose energy-dependent cuts (i.e. in $N_{\rm phot}$ bins) using pre-defined {\gray} efficiencies %, see fig. \ref{BDTefficiency} 
(e.g. being more restrictive at low energies where the background is more difficult to eliminate but the signal {\gray}s are more abundant).

%\begin{figure}
%  \includegraphics[width=4cm]{Figure_BDTefficiencies}
%  \caption{{\gray} efficiency required of the BDT cut, and resulting background in increasing bins in $N_{\rm phot}$}
%  \label{BDTefficiency}
%\end{figure}

Note that we explicitly chose not to use goodness-of-fit type discriminants, as these require an excellent understanding of detector characteristics for proper application.  %See \citep{ParisMVA} and references therein for a full description of Paris-MVA as applied to H.E.S.S.

\subsection{Energy Reconstruction (``Oak'')}

Paris-MVA uses a simple, fast energy-reconstruction procedure, baptized ``Oak''.  Monte-Carlo simulations provide charge profiles as a function of impact parameter ($q$ vs. $R$), for given ranges in Energy $E$, and in depth of Shower Maximum $H$.
%For an operational detector, the response as a function of Zenith angle, detector efficiency (mirror/cone reflectivity, collection efficiency, etc.), and source offset in the camera are also taken into account, but for this work the Monte-Carlo simulations currently have a single value for each.  
The simulated profiles obtained are first inverted, to obtain profiles of $E$ vs. $q$ for given ranges in $R$; these profiles are stored for the reconstruction phase.

In the energy-reconstruction phase, for a given telescope, 
the reconstructed $H$ and $R$ (determined from the Hillas parameters) allow to choose which histograms in $R$ and which bins in $q$ 
enter into the interpolation (using a weighted tree) for the energy $E_{\rm tel}$ estimated based on that telescope's information.  
In the final step, the energies of all the telescopes implicated in an event are combined in a sum weighted by each telescope's charge, 
giving the estimate of the {\gray} energy.

For Paris-MVA in CTA, a linear correction with $\log(E)$ (determined from the Monte-Carlo simulations) is made in order to take out residual biases, 
for comparative plots with other analysis techniques.
%Note that for operational detectors, such biases are automatically corrected by standard forward-folding spectral determination techniques,
%which thus permit to plot fitted spectra versus the true photon energy. 
Note that for operational detectors, the standard spectral determination techniques take into account and correct for these biases, 
allowing to plot fitted spectra versus the true photon energy. 
%  The resulting energy-dependent resolution and bias shown in fig. \ref{MVA_ResBias}, with 
The results for example at $1\;{\rm TeV}$ show an energy resolution of 13\% with $\sim 0$\% bias being achieved.

%%Comment J.Rico: About "Oak":
%It is mentioned in the last paragraph that a correction for reducing the bias is done:
%  - not clear what the "comparative plots" refers to
%  - next sentense seems to imply that a working detector corrects the bias by itself (automatically). I think this is not the case, the methods for the unfolding of the spectra need to be implemented once the response of the detector is known well enough, right? Maybe there is something else here I am not aware of? in that case it would probably worth a reference
%  - Also, the unfolding of the detector response (bias+resolution) can be done with forward or "backward" unfonding techniques.
%% Reply:
% It refers to the comparative plots with other analysis techniques, so added "with other analysis techniques."
% Okay, "automatically" may be misread.  To take into account also the comment on backward unfolding techniques, the sentence has been simplified to 
% ... for operational detectors, the standard spectral determination techniques take into account and correct for these biases, allowing ..."

%\begin{figure}
%  \includegraphics[width=10cm]{Figure_MVA_ResBias}
%  \caption{Energy resolution and bias given by the Oak procedure as a function of $\log_{10}({\rm reconstructed energy})$.}
%  \label{MVA_ResBias}
%\end{figure}

\subsection{%Standard and 
Newly-Developed Parameters}

%Two standard ``Hillas'' moment-based parameters image parameters are used in Paris-MVA: Mean-scaled width and Mean-scaled length (MSCW \& MSCL).

%The 3D-Model \citep{model3D} provides three further event parameters for input to Paris-MVA. The 3D-model models the shower as a 3D-photosphere with the form of a prolate ellipsoid (Rugby-ball shape), characterized by its 3D-Reduced-Length, 3D-Reduced-Width (``Reduced'' implying corrected for Zenith-angle dependence), its Centroid corresponding to the Depth of Shower Maximum and its number of Cherenkov photons $N_{\rm phot}$.  A maximum-likelihood fit is carried out to reproduce the model's charge expectations on the pixels in the cameras.
%The Paris-MVA analysis uses 3D-Reduced-Width, its error as by the likelihood maximization, and the fitted Depth of Shower Maximum, as three further powerful discriminant parameters.
%, see fig. \ref{NewParamDist} for an illustration of their distributions in one sample training bin.  
%Paris-MVA also uses $N_{\rm phot}$ as a simple proxy for the energy, in order to place the event in one of the BDT training bins used.

The three newly-developed parameters input to the BDT, described in \citep{ParisMVA}, are based on 
the exploitation of the differences in shower development for signal {\gray}s versus the hadronic background.
These give, for example, incoherences in the 3D-model (mis-)fit of a hadronic shower with a gamma-ray model.
%A {\gray} shower has a clean and symmetric development, whereas a proton or hadron shower is broken-up and asymmetric,
%so for example we can take advantage of the azimuthal symmetry of the gamma ray shower versus the asymmetry of the proton shower.
%For the 3D-model, the (mis-)fit of a hadronic shower with a gamma-ray model gives incoherences that can be exploited.  
The images (pixel values) predicted in each camera by the 3D-model minimization %(see fig. \ref{SimVsModel}) 
can be used as the basis for a new set of Hillas parameters (\emph{HillasOnModel}) which are then used to define the new discriminant parameters.

%\begin{figure}
%  \includegraphics[width=6cm]{Figure_SimVsModel}
%  \caption{Illustration from HESS data and simulations of a simulated {\gray} versus its 3D-fitted model, compared with a real hadron and its model.}
%  \label{SimVsModel}
%\end{figure}

The first new discriminant parameter, $\Omega$, is defined as the angle between the shower direction determined from the original \emph{Hillas} parameters, and that determined from the \emph{HillasOnModel} parameters.
Two further sensitive discriminant parameters can be defined, based on the Oak energy reconstruction algorithm. 
By reversing the \emph{Oak} procedure, given the energy determined from Oak, we can ``predict'' the charge seen in each telescope, so giving $\Delta Q$, the sum of the squared difference between the observed and predicted charges, divided by the summed observed charges.
Finally, we can define $R_E$: the ratio of the energy as determined from the initial Hillas parameters to that found from HillasOnModel.
These three new parameters %(see distributions of signal vs. background for an example training bin shown in fig. \ref{NewParamDist}), 
have been shown for H.E.S.S.\ to provide a gain of 20\% in background rejection for fixed {\gray} efficiency.

%Comment J.Rico: In last sentence of section "Newly-Developed Parameters", could a reference be added?
%Reply: this is the "ParisMVA" paper reference, so to be clear we've added this to the 1st sentence:
%"The three newly-developed parameters input to the BDT, described in \citep{ParisMVA}, are based on ..."

%\begin{figure}
%  \includegraphics[width=12cm]{Figure_MVA_NewParamDist}
%  \caption{Distributions of the (log10 of) the newly-defined Paris-MVA discriminant parameters on CTA simulations in one example training bin}
%  \label{NewParamDist}
%\end{figure}

\section{Adaptation for CTA, Simulations, \& BDT Training}

Some slight modifications of the 3D-model fitting procedure have been implemented.  For the initial condition to the fitting procedure, we use length, width, and shower maximum estimations based only on largest contiguous cluster in each telescope (after standard image cleaning).  Also, the information from a telescope is used in the 3D-fit only if it has a total charge $>30\,{\rm photoelectrons}$ and with at most 3 clusters of contiguous pixels after standard image cleaning.
%These modifications improve the fit convergence for the largest, high-multiplicity events.
These modifications allow to adapt the fit convergence for the largest, high-multiplicity events seen with CTA.

%Comment J.Rico: First paragraph of section "Aaptation for CTA...". Could the last sentence (improvement seen wrt HESS analysis) be quantified?
%Reply: Sorry, we should clarify.  The improvement is with respect to the "blind" application of the method to CTA, prior to these modifications.  
% Without modifications, the method is not well-adapted for the largest, high-multiplicity events.  So, to clarify, the sentence is changed to 
% "These modifications allow to adapt the fit convergence for the largest, high-multiplicity events seen with CTA."

All the results were obtained using simulations from the SimTelArray package \citep{StdMPIK}, 
at $20^\circ$ from Zenith, with 
  {\gray}s simulated  on-axis 
  between $3\,{\rm GeV}$--$300\,{\rm TeV}$, 
  diffuse electrons in the same energy range, and 
  diffuse protons between $5\,{\rm GeV}$--$500\,{\rm TeV}$.

\section{Results for CTA}

\begin{figure}
  \includegraphics[width=16.cm]{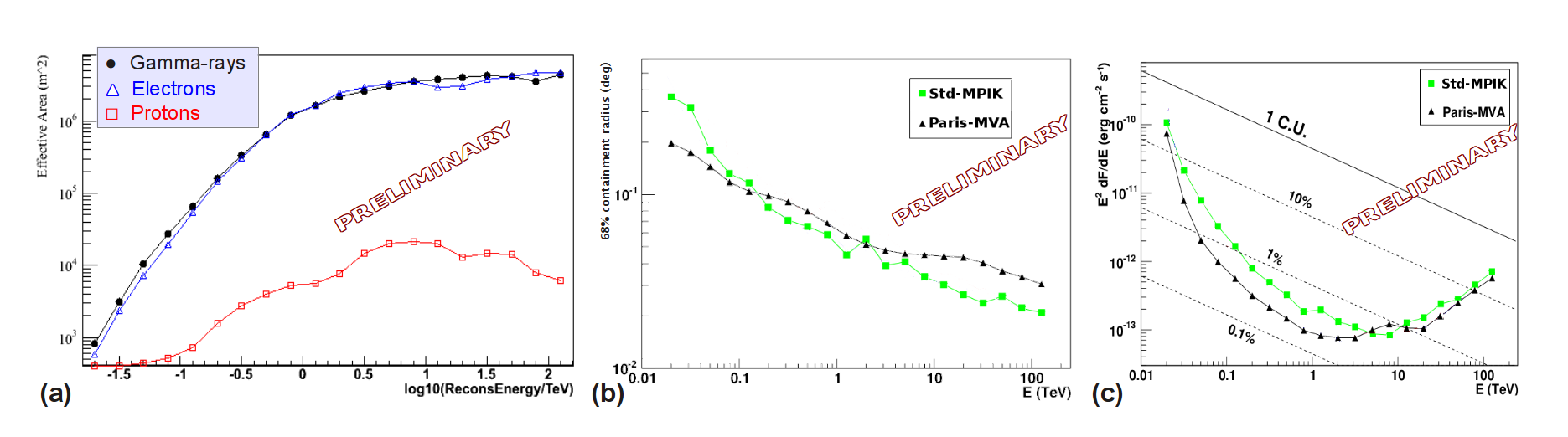}
  \caption{(a) Effective collection areas after Paris-MVA and energy-dependent point source $\theta^2$ cut; 
	  (b) Angular resolution (68\% containment radius) as function of Energy, for point source {\gray}s after Paris-MVA cuts, compared with the standard analysis.
	  (c) Differential sensitivity for a point-source as calculated for array ``I'', for Paris-MVA  vs.\ standard analysis.}
  \label{MVA_CollA_AngRes_DiffSensi}
\end{figure}

The effective area after Paris-MVA BDT cuts and energy-dependent $\theta^2$ cut is shown in fig. \ref{MVA_CollA_AngRes_DiffSensi}(a), for {\gray}s, and for the electron and proton background.  For the point-source {\gray}s, the angular resolution (68\% containment radius) as a function of energy is shown in fig. \ref{MVA_CollA_AngRes_DiffSensi}(b).

%\begin{figure}
%  \includegraphics[width=14cm]{Figure_MVA_CollA_AngRes}
%  \caption{(a) Effective collection areas after Paris-MVA and energy-dependent point source $\theta^2$ cut; 
%	  (b) Angular resolution (68\% containment radius) as function of Energy, for point source {\gray}s after Paris-MVA cuts, compared with the standard analysis.}
%  \label{MVA_CollA_AngRes}
%\end{figure}

For the sensitivity calculation shown here, the proton background is scaled by a factor of 1.1 to take into account the presence of heavier species in the charged cosmic-ray flux, and proton and electron background spectra are taken into account.
The performance of CTA is shown in fig. \ref{MVA_CollA_AngRes_DiffSensi}(c), as applied to the example array ``I'' (see \cite{DesignConcepts}), an array consisting of 77 telescopes (3 large size telescopes, 24m diameter, $5^\circ$ FoV; 18 mid-size, $\sim 12$m, $8^\circ$; and 56 small-size 7.4m, $10^\circ$), which is optimized for a good wide-energy-range performance.  The figure shows the differential point-source sensitivity in energy bins (5 bins per decade, requirement of $5\sigma$ per bin) for 50h observation, for the Paris-MVA analysis as compared with that derived from a standard Hillas analysis (see \cite{DesignConcepts}).

%\begin{figure}
%  \begin{minipage}[c]{0.6\textwidth}
%    \includegraphics[width=8cm]{Figure_MVA_DiffSensi}
%  \end{minipage}
%  \hfill
%  \vspace -1cm
%  \begin{minipage}[c]{0.3\textwidth}
%    \caption{Differential sensitivity for a point-source as calculated for array ``I'', for Paris-MVA  vs.\ standard analysis.}
%    \label{MVA_DiffSensi}
%  \end{minipage}
%\end{figure}

%\begin{figure}
%  \includegraphics[width=6cm]{Figure_MVA_DiffSensi}
%  \caption{Differential sensitivity for a point-source as calculated for array ``I'', for Paris-MVA  vs.\ standard analysis.}
%  \label{MVA_DiffSensi}
%\end{figure}

%\begin{figure}
%\floatbox[{\capbeside\thisfloatsetup{capbesideposition={right,top},capbesidewidth=5cm}}]{figure}[\FBwidth]
%{\caption{Differential sensitivity for a point-source as calculated for array ``I'', for Paris-MVA  vs.\ standard analysis.}\label{MVA_DiffSensi}}
%{\includegraphics[width=8cm]{Figure_MVA_DiffSensi}}
%\end{figure}

\section{Conclusions}

%Thanks to the conception of Paris-MVA in a flexible manner and its implementation within a standard TMVA environment, this enabled it to be rather quickly adapted from H.E.S.S.\ to be applied to CTA.

Paris-MVA was conceived to be flexible, and was implemented within a standard TMVA environment; these features enabled it to be rather quickly adapted from H.E.S.S.\ to be applied to CTA.

%Comment J.Rico: Conclusions: some word(s) missing between "made" and "initialization"  
% Yes, thanks, added "made _to the_ initialization"

Some modifications have been made to the initialization of the fit in the 3D-model fitting procedure to achieve reasonable convergence at the highest energies (difficult to reach with H.E.S.S.\ with high statistics), though further work still needs to be done, perhaps to the 3D-model itself (e.g., to take into account the instrument's integration window).

It is readily apparent that the application of the advanced Paris-MVA analysis technique, combining within the MVA framework both previously-known and newly-defined sensitive discrimination variables, allows the performance of the CTA observatory to be improved by a factor $\sim 2$ over the standard analysis.

Comparative studies of the performance under different array layouts and with the different electronics concepts have begun, and should allow the future CTA array to be optimized \citep{CTA_Design} taking into account such advanced analyses.


\begin{thebibliography}{9}

\bibitem{DesignConcepts}
The CTA Consortium, ``CTA Design Concepts'',  \emph{Exp. Astron.} 32 (2011) 193--316

\bibitem{model3D}
M. Lemoine-Goumard, B. Degrange, M. Tluczykont, \emph{Astropart.Phys.} 25 (2006) 195--211

\bibitem{TMVA}
\url{http://tmva.sourceforge.net/}

\bibitem{ParisMVA}
Y. Becherini et al., \emph{Astropart. Phys. 34} 12 (2011) 858--870

\bibitem{StdMPIK}
K. Bernl\"ohr, \emph{Astropart. Phys.} 30 (2008) 149

%\bibitem{Root}
%\url{http://root.cern.ch/}

\bibitem{CTA_Design}
V. Stamatescu, et al. ``Towards an optimized design for Cherenkov Telescope Array'', \emph{this conference}

\end{thebibliography}
\end{document}